# Ptychographic lens-less polarization microscopy

**Jeongsoo Kim,[1] Seungri Song,[1] Bora Kim,[2] Mirae Park,[3] Seung Jae Oh,[3,4] Daesuk Kim,[5] Barry Cense,[1,6] Yong-Min Huh,[3,4,7] Joo Yong Lee,[2] and Chulmin Joo[1,*]**

[1]*Department of Mechanical Engineering, Yonsei University, Seoul 03722, Republic of Korea*
[2]*Department of Ophthalmology, Asan Medical Center, University of Ulsan College of Medicine, Seoul 05505, Republic of Korea*
[3]*Department of Radiology, College of Medicine, Yonsei University, Seoul 03722, Republic of Korea*
[4]*YUHS-KRIBB Medical Convergence Research Institute, Seoul 03722, Republic of Korea*
[5]*Department of Mechanical System Engineering, Jeonbuk National University, Jeonju 54896, Republic of Korea*
[6]*Optical and Biomedical Engineering Laboratory, Department of Electrical, Electronic & Computer Engineering, The University of Western Australia, Perth, WA 6009, Australia*
[7]*Department of Biochemistry & Molecular Biology, College of Medicine, Yonsei University, Seoul 03722, Republic of Korea*
*\*cjoo@yonsei.ac.kr*

**Abstract:** Birefringence, an inherent characteristic of optically anisotropic materials, is widely utilized in various imaging applications ranging from material characterizations to clinical diagnosis. Polarized light microscopy enables high-resolution, high-contrast imaging of optically anisotropic specimens, but it is associated with mechanical rotations of polarizer/analyzer and relatively complex optical designs. Here, we present a novel form of polarization-sensitive microscopy capable of birefringence imaging of transparent objects without an optical lens and any moving parts. Our method exploits an optical mask-modulated polarization image sensor and single-input-state LED illumination design to obtain complex and birefringence images of the object via ptychographic phase retrieval. Using a camera with a pixel resolution of 3.45 μm, the method achieves birefringence imaging with a half-pitch resolution of 2.46 μm over a 59.74 mm$^2$ field-of-view, which corresponds to a space-bandwidth product of 9.9 megapixels. We demonstrate the high-resolution, large-area birefringence imaging capability of our method by presenting the birefringence images of various anisotropic objects, including a birefringent resolution target, liquid crystal polymer depolarizer, monosodium urate crystal, and excised mouse eye and heart tissues.

## 1. Introduction

Since the invention of the first compound microscope, several hardware improvements, including optical components and image sensors, have been made over the years. However, its basic design consisting of multiple lenses to form an optical image of an object has not changed significantly. This conventional imaging system is beset by the inherent trade-offs between the spatial resolution and field-of-view (FoV), limiting their space–bandwidth product (SBP), which can be calculated as the FoV divided by the square of the spatial resolution [1]. In other words, one can only image the fine details of an object with high resolution in a small region. Because imaging systems require a large SBP to handle multiple applications, the conventional imaging design relies on complicated optical and mechanical architectures to increase the SBP, resulting in a bulky and expensive imaging setup.

In recent years, as advanced computational algorithms have been developed and computational resources became more powerful, various computational microscopy techniques were developed to overcome the SBP limitation of conventional imaging systems [2-5]. Lens-less microscopy is one of the representative computational imaging platforms that were developed to address the SBP limitation, while enabling a "lean" optical architecture [6-9]. Because the final image can be obtained with unit magnification by placing the object directly on the image sensor, the imaging FoV is only limited by the size of the image sensor, typically a few millimeter, and the resolution is determined by the pixel size of the imaging sensor [9]. Moreover, various pixel-super resolution methods can be employed to further enhance the resolution, using measurement diversities based on multiple object heights [10,11], object/sensor translation [12,13], multiple wavelengths [14,15], etc. In addition, synthetic aperture-based lens-less imaging technology enables high-resolution imaging by enlarging the effective numerical aperture (NA) using angle varied illumination [16]. The lens-less microscope can also retrieve the complex information of the object from intensity-only measurements via back projection and a phase retrieval algorithm. Taking advantage of these features, namely high information throughput, small form-factor, and cost-effectiveness, various imaging modalities such as phase and fluorescence imaging have been demonstrated in a lens-less platform [17,18], and numerous applications have been proposed, including cell observation [19], disease diagnosis [6], and air quality monitoring [20].

Birefringence is the polarization-dependent refractive index of a material, and it is the intrinsic optical property of optically anisotropic materials [21,22]. The anisotropy of the refractive index arises from the ordered arrangement of microstructures within the material, and birefringence can thus be used to effectively characterize the structural details of the material. Birefringence has been broadly studied for various applications such as material inspection [23-25] and biomedical diagnosis [26-28]. Polarization light microscopy (PLM) is conventionally used to measure anisotropy properties (i.e., phase retardation and optic-axis orientation) of the transparent materials. Its operation involves multiple image acquisitions with mechanical rotations of the polarizer/analyzer in a conventional optical microscope, and thus is associated with a relatively complex optical setup. In addition, PLM, as in conventional microscopes, has limited SBP. In order to circumvent these limitations, birefringence imaging has recently been demonstrated on various imaging platforms, such as digital holographic microscopy [29], ptychography [30], single-pixel imaging [31], differential phase-contrast microscopy [32], and lens-less holographic microscopy [33,34]. Among them, lens-free holographic polarization microscopy enables large-area birefringence imaging in a lens-free manner, but two sets of raw holograms must be taken with illuminations in two different polarization states, which requires precise image alignment. Recently, this imaging method has been integrated with deep learning to achieve single-shot birefringence imaging, albeit requiring training with a large number of datasets [35].

Here, we present a novel form of polarization-sensitive microscopy, which allows for inertia-free, large-area, high-resolution birefringence imaging without any lenses. Our polarization-sensitive ptychographic lens-less microscope (PS-PtychoLM) method adopts a high-resolution, large-FoV imaging capability of a mask-modulated ptychographic imager [36] while quantitatively measuring the birefringence properties of transparent samples with a single-input-state illumination and polarization-diverse imaging system. Compared to conventional polarization imaging techniques, our PS-PtychoLM does not involve any mechanical rotation of the polarizer/analyzer and translation of object or light source, which makes the system more robust and easier to operate. A Jones-matrix analysis is used to formulate the PS-PtychoLM image reconstruction process. We demonstrate high-accuracy polarization-sensitive (PS) imaging capability over a large-FoV by presenting the birefringence images of various anisotropic objects, including a birefringent resolution target, liquid crystal polymer depolarizer, monosodium urate (MSU) crystal, and mouse eye and heart tissue sections.

## 2. Methods

*2.1 Optical setup*

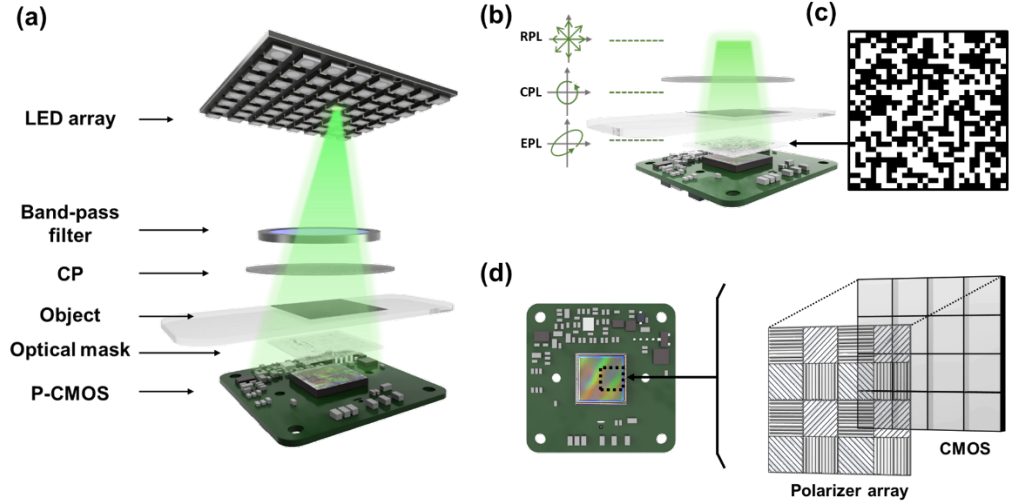

Fig. 1. (a) Schematic of our polarization-sensitive ptychographic lens-less microscope (PS-PtychoLM). A programmable light-emitting diode (LED) array is turned on sequentially to generate angle-varied illumination. (b) Unpolarized LED light passes through a band-pass filter and circular polarizer (CP) to be circularly polarized and to illuminate an object. (c,d) The light transmitted through the object is intensity modulated by an optical mask, and then recorded through four different polarization channels of a polarization camera (P-CMOS) (RPL: randomly polarized light, CPL: circularly polarized light, EPL: elliptically polarized light).

Our PS-PtychoLM was built on a mask-modulated ptychographic imager [36], which provides inertia-free complex image reconstruction with a programmable LED array. A schematic of the PS-PtychoLM design is depicted in Figure 1. We employed a custom-built programmable LED array for angle-varied illumination. The LED array was composed of high-power LEDs (Shenzhen LED color, APA102-202 Super-LED, center wavelength = 527 nm, full-width at half-maximum bandwidth ~ 30 nm), and featured a 4 mm pitch with 15 × 15 elements that can be controlled independently. Only the central 9 × 9 elements of the LED array were used, and a total of 81 raw intensity images were captured for each LED illumination. The distance from the LED array to the object was set to be ~400 mm, providing illumination angles of -2.29 to 2.29°. Prior to illuminating an object, the LED light passed through a band-pass filter (Thorlabs, FL514.5-10, center wavelength = 514.5 nm, full-width at half-maximum bandwidth = 10 nm) to increase temporal coherence, after which the light was circularly polarized using a zero-order circular polarizer (Edmund optics, CP42HE). If the object is optically anisotropic, the light passing through the object becomes elliptically polarized (Figure 1b). The transmitted light was then intensity modulated by a binary amplitude mask (Figure 1c). The mask was fabricated by coating a 1.5 mm thick soda-lime glass slab with chromium metal with a random binary pattern that was employed in Ref. 36. The transparent-to-light-blocking area ratio of the optical mask was designed to be one and the pattern had a feature size of 27.6 μm. The modulated light was then captured using a board-level polarization-sensitive complementary metal-oxide-semiconductor image sensor (P-CMOS, Lucid Vision Labs, PHX050S1-PNL) with a pixel resolution of 2048 × 2448 and a pixel size of 3.45 μm, allowing for a FoV of 7.07 mm x 8.45 mm. The camera was equipped with four different directional polarization filters (0°, 90°, 45°, and 135°) on the image sensor for every four pixels, allowing the information of each polarization state to be captured simultaneously (Figure 1d). PS-PtychoLM could thus acquire polarization images along a specific orientation required for computing the birefringence distribution without mechanical rotation of the polarizers or variable retarder.

*2.2 Jones-matrix analysis for PS-PtychoLM birefringence imaging*

The Jones-matrix analysis was used to formulate the PS-PtychoLM image reconstruction process, in which the polarization status of light is represented by a Jones vector, and the optical elements, including specimens, are represented by Jones matrices [37,38]. Light field output ($E_{out,\psi}$) measured at each polarization channel through our imaging system can be represented as:

$$E_{out,\psi} = J_{d,\psi} J_s E_{in}, \tag{1}$$

where $E_{in}$ denotes the Jones vector for the incident light of intensity I with $E_{in} = \sqrt{I/2}[1 \quad i]^T$, $J_s$ denotes the Jones matrix for nondepolarizing thin specimen, $J_{d,\psi}$ denotes the Jones matrix for the linear polarizer on the detector along $\psi$ direction ($\psi = 0°, 45°, 90°,$ and $135°$). Jones matrix for a thin sample ($J_s$) can be represented with a phase retardance magnitude ($\delta$) and an optic-axis orientation ($\theta$) as:

$$J_s = RDPR^{-1}, \tag{2}$$

where $R = [cos\theta \quad -sin\theta; sin\theta \quad cos\theta]$ is a rotation matrix defined by optic-axis orientation $\theta$, $P = [e^{i\delta/2} \quad 0; 0 \quad e^{-i\delta/2}]$ is the phase retardation matrix, and $D$ is the diattenuation matrix which is negligible in thin specimens. As a result, the Jones matrix for thin specimens can be formulated as:

$$J_s = e^{\frac{i\delta}{2}} \begin{bmatrix} cos^2\theta + e^{-i\delta}sin^2\theta & (1-e^{-i\delta})sin\theta cos\theta \\ (1-e^{-i\delta})sin\theta cos\theta & sin^2\theta + e^{-i\delta}cos^2\theta \end{bmatrix}. \tag{3}$$

The Jones matrix for four different polarization channels on the detector along $\psi$ direction is expressed as:

$$J_{d,\psi} = \begin{bmatrix} cos^2\psi & sin\psi cos\psi \\ sin\psi cos\psi & sin^2\psi \end{bmatrix}. \tag{4}$$

The measured intensity values for each polarization channel ($I_\psi$) can be written as:

$$I_\psi \propto |E_{out,\psi}|^2 = E_{out,\psi} E_{out,\psi}^*. \tag{5}$$

Using Eq. (1-5), one can easily obtain the expressions for the light intensity measured along the four polarization channels as follows:

$$I_0 = \frac{I}{2}(1 - sin\delta sin2\theta), \tag{6}$$

$$I_{45} = \frac{I}{2}(1 + sin\delta cos2\theta), \tag{7}$$

$$I_{90} = \frac{I}{2}(1 + sin\delta sin2\theta), \tag{8}$$

$$I_{135} = \frac{I}{2}(1 - sin\delta cos2\theta). \tag{9}$$

Then, we combine Eq. (6-9) to introduce two auxiliary quantities $Q_1$ and $Q_2$ defined as:

$$Q_1 = \frac{I_{90} - I_0}{I_{90} + I_0} = sin\delta sin2\theta, \tag{10}$$

$$Q_2 = \frac{I_{45} - I_{135}}{I_{45} + I_{135}} = sin\delta cos2\theta. \tag{11}$$

Finally, the spatial distribution of retardance magnitude and the optic-axis orientation of the birefringent sample can be reconstructed as:

$$\delta(x,y) = \sin^{-1}\sqrt{Q_1^2 + Q_2^2}, \quad (12)$$

$$\theta(x,y) = \frac{1}{2}\tan^{-1}\left(\frac{Q_1}{Q_2}\right). \quad (13)$$

*2.3 PS-PtychoLM forward model*

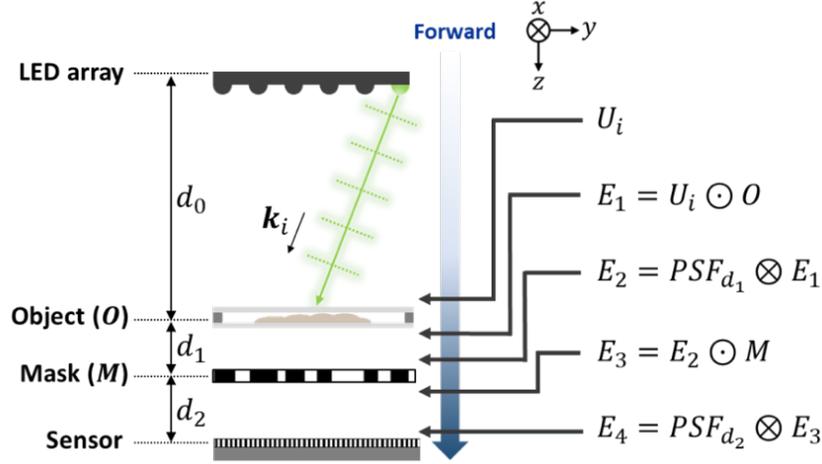

Fig.2. PS-PtychoLM image formation. $U_i$, $O$, and $M$ represent the incident light field from the $i^{th}$ LED element, complex information of the object, and optical mask, respectively. $d_0$, $d_1$, and $d_2$ denote the distances between the LED array and object, the object to the mask, and the mask to sensor plane, respectively. The spatial coordinate $r$ is omitted for simplicity.

The forward imaging model of PS-PtychoLM was used to predict the manner in which the optical wave passing through the object and mask was recorded at the image sensor (Figure 2). Consider an object illuminated by the $i^{th}$ element in the LED array. In the implementation, the distance between the LED elements and the object was large (~400mm), and thus the illumination light wave from the $i^{th}$ LED element ($U_i$) can be assumed to be a plane wave in the object plane, which is expressed as:

$$U_i(r) = exp[jk_i \cdot r], \quad (14)$$

where $r = (x, y)$ denotes the spatial coordinate, $j$ is the imaginary unit, $k_i = (k_{x,i}, k_{y,i})$ is the incident wave vector, and · is the inner product of vectors. Using the PS-PtychoLM configuration (Figure 2), the incident wave can be re-written as $k_{x,i} = -ksin[tan^{-1}(x_{LED,i}/d_0)]$, $k_{y,i} = -ksin[tan^{-1}(y_{LED,i}/d_0)]$, where $k$ is the wave number of light and $(x_{LED,i}, y_{LED,i})$ is the lateral location of the $i^{th}$ LED element. The illumination field then interacts with the object and the exit light field in the object plane can be written as $E_1(r) = U_i(r) \odot O(r)$, where $O(r)$ is the object complex information and $\odot$ stands for the element-wise multiplication. The exit wave $E_1$ then propagates to the mask plane and the wave incident on the mask is given by $E_2(r) = PSF_{d_1} \otimes E_1(r)$, where $PSF_d$ represents the point spread function (PSF) for free-space propagation over the propagation distance $d$, and $\otimes$ denotes a convolution operation. In the implementation, free-space propagation was operated in the Fourier space using the angular spectrum method [39,40]. The light field $E_2$ is modulated by the optical mask, resulting in $E_3(r) = E_2(r) \odot M(r)$, where $M(r)$ is the mask transmission function. The modulated light field $E_3$ further propagates to the sensor plane to result in $E_4(r) = PSF_{d_2} \otimes E_3(r)$. The intensity image captured by the image sensor with the $i^{th}$ LED illumination ($I_i$) can then be written as:

$$I_i(r) = |E_4(r)|^2 = |PSF_{d_2} \otimes [PSF_{d_1} \otimes \{U_i(r) \odot O(r)\} \odot M(r)]|^2. \tag{15}$$

## 2.4. PS-PtychoLM image reconstruction procedure

By sequentially turning on each LED, a set of intensity images $I_i$ ($i = 1, 2, ..., L$) is acquired and used for PS-PtychoLM image reconstruction. In our demonstration, we recorded 81 intensity images with a size of $M \times M$ by turning on the 81 LEDs centered at the optical axis. In order to reconstruct complex information of object and binary mask, we used the regularized ptychographic iterative engine (rPIE) [41] and the reciprocal-space up-sampling method [42]. A detailed reconstruction procedure is presented as follows.

Step 1) We first initialize the object and mask as constant all-one matrices with a size of $(N \times M) \times (N \times M)$, where $N$ represents the up-sampling ratio. Under the $i^{th}$ LED illumination, the field at the sensor plane ($E_4$) can be estimated as Eq. (14)-(15) using the PS-PtychoLM forward model. Since the recorded intensity image ($I_i$) has a size $M \times M$, the estimated $E_4$ is down-sampled with a factor of $N$, then the information is updated with the acquired intensity image and then up-sampled again by $N$-times. This process is expressed as follows:

$$E_4' = E_4(r) \odot \uparrow_N \left\{ \sqrt{\frac{I_i}{\downarrow_N [|E_4(r)|^2 \otimes ones(N \times N)]}} \right\}, \tag{16}$$

where $\uparrow_N()$ and $\downarrow_N()$ denote the $N$-times nearest-neighbor up-sampling and down-sampling operations, respectively. $ones(N \times N)$ represents an $N \times N$ all-one matrix. The convolution procedure in the denominator of Eq. (16) using $ones(N \times N)$ enforces the intensity sum of each $N \times N$ segment of the result to be equal to the corresponding pixel of the acquired image [42]. Note that, in our implementation, the $N \times N$ all-one matrix was zero-padded to the size $(N \times M) \times (N \times M)$ and then the convolution operation was performed in the Fourier domain. The value of $N$ was set to 4.

Step 2) The updated light field $E_4'$ is then back-propagated to the mask plane as:

$$E_3'(r) = conj(PSF_{d_2}) \otimes E_4'(r), \tag{17}$$

where $conj()$ represents complex conjugation.

Step 3) Complex information of the mask and light field above the mask is updated using the rPIE algorithm as:

$$M'(r) = M(r) + \beta \cdot \frac{conj[E_2(r)]}{(1-\alpha)[|E_2(r)|^2] + \alpha \cdot \max_r[|E_2(r)|^2]} \odot [E_3'(r) - E_3(r)], \tag{18}$$

$$E_2'(r) = E_2(r) + \beta \cdot \frac{conj[M'(r)]}{(1-\alpha)[|M'(r)|^2] + \alpha \cdot \max_r[|M'(r)|^2]} \odot [E_3'(r) - E_3(r)], \tag{19}$$

where $\alpha$ and $\beta$ are the weighting parameters of the rPIE algorithm. We set $\alpha = 0.8$ and $\beta = 0.5$ in the implementation.

Step 4) The updated light field $E_2'$ is then back-propagated to the object plane:

$$E_1'(r) = conj(PSF_{d_1}) \otimes E_2'(r). \tag{20}$$

Step 5) Lastly, the complex information of object is updated with the light field $E_1'$ as:

$$O'(r) = O(r) + \beta \cdot \frac{conj[U_i(r)]}{(1-\alpha)[|U_i(r)|^2] + \alpha \cdot \max_r[|U_i(r)|^2]} \odot [E_1'(r) - E_1(r)]. \tag{21}$$

The object $O$ and mask $M$ are replaced with the updated object $O'$ and mask $M'$ before starting the update procedures (steps 1-5) using the LEDs in different positions. One iteration is complete if the update procedure is performed for all the LEDs. When the object and mask converge, the reconstruction is finished, and in our experiment, it generally took approximately 10 iterations to complete.

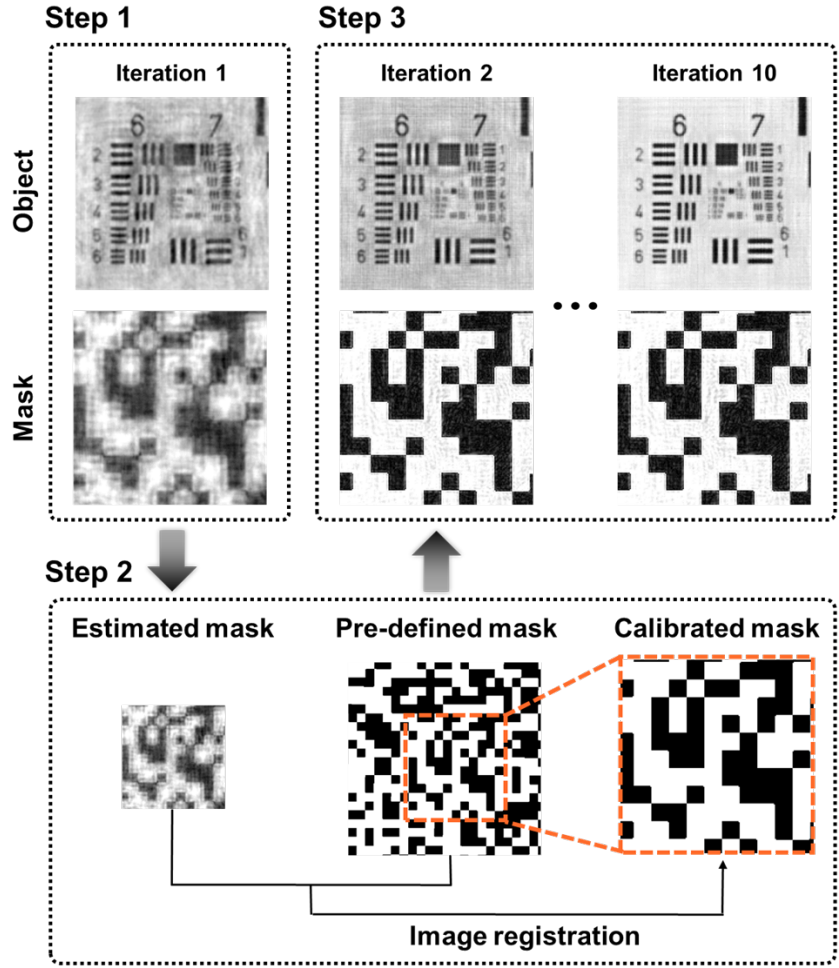

Fig. 3. Procedures to obtain the actual mask pattern in the PS-PtychoLM image reconstruction.

In our PS-PtychoLM implementation, a binary mask with a pre-defined pattern was employed for robust image reconstruction [36]. Even though the mask pattern is pre-defined and fabricated with high fidelity, it is extremely challenging to align the mask in a designated space in the optical setup. In practice, the actual mask pattern that contributes to the image formation may have a lateral shift and rotation, and it may be a small portion of the designed mask pattern. This uncertainty on the actual mask pattern may result in a complete failure of image recovery. We obtained the mask pattern that contributed to the image formation with the three steps outlined below (Figure 3):

Step 1) We first perform a blind recovery using a guessed mask pattern. The objective of this blind recovery is to obtain a rough estimate of the mask pattern. In this step, the complex object information and the mask pattern are initialized by all-one matrices, respectively. Then, one iteration of image recovery is conducted, and then rough estimates of the mask pattern and the object are obtained.

Step 2) We then conduct cross-correlation of the estimated mask pattern obtained from Step 1 and the designed mask pattern to determine the amount of lateral shift and rotation of the actual mask with respect to the designed mask pattern. The resultant information was used to further refine the actual mask pattern that contributes to the image formation.

Step 3) The obtained actual mask pattern is then used as the initial guess for the subsequent iteration of image recovery.

## 3. Results

*3.1 Numerical simulation*

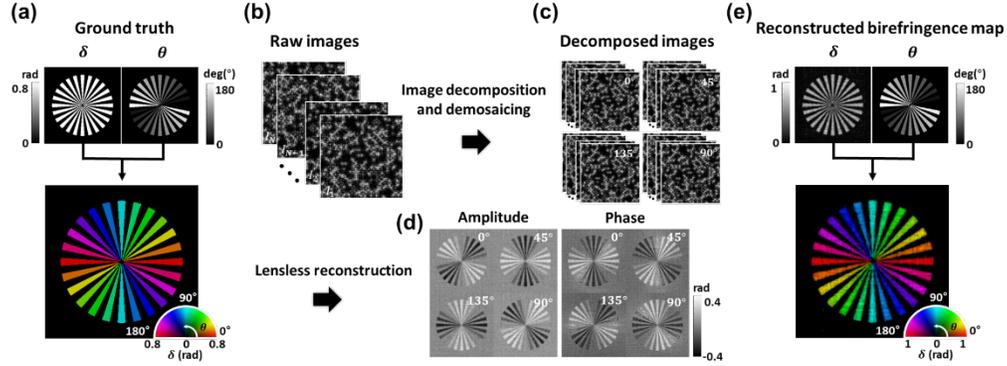

Fig. 4. PS-PtychoLM birefringence map reconstruction procedure and simulation results. (a) Ground truth birefringence information of a simulation target. (b,c) PS-PtychoLM captures raw images under angle-varied LED illuminations, and each recorded image is decomposed into the images in four polarization channels. The missing pixel information in each image is interpolated using a polarization demosaicing scheme. (d) The amplitude and phase information of each polarization channel are then reconstructed using a ptychographic reconstruction algorithm, and further used to obtain phase retardation and optic-axis orientation maps of the object based on Jones-matrix analysis. (e) Obtained birefringence information can be jointly presented using a pseudo-color map that encodes retardation and optic-axis orientation with intensity and color, respectively.

Figure 4 outlines the PS-PtychoLM birefringence map reconstruction procedure and numerical simulation results of a simulated birefringence object. We considered a Siemens birefringent object, of which each sector form was characterized by a phase retardation of $\pi/4$ and their optic axes oriented along the longer sides. The corresponding birefringence map is presented by a pseudo-color map to denote optic-axis orientation ($\theta$) and retardation ($\delta$) with color and intensity, respectively (Figure 4a). The PS-PtychoLM measurements were numerically performed (Figure 4b). The recorded images by the P-CMOS under angle-varied LED illuminations were first decomposed into four images at different polarization channels. The resultant images were then processed for polarization demosaicing to estimate missing polarization information at the neighboring pixels (Figure 4c). For the demosaicing, we used Newton's polynomial interpolation model [43] because it was found to recover images with high fidelity in both low- and high-frequency features, compared to other schemes (Appendix A). The amplitude and phase images of each polarization channel were then reconstructed using a ptychographic reconstruction algorithm [41] (Sec. 2.3 and 2.4) and further used to obtain phase retardation and optic-axis orientation of the birefringent object based on the Jones-matrix analysis (Figure 4d). The reconstructed birefringence information is presented in Figure 4e, which agrees with the ground truth map (Figure 4a).

## 3.2 Experimental demonstration

### 3.2.1. Isotropic targets

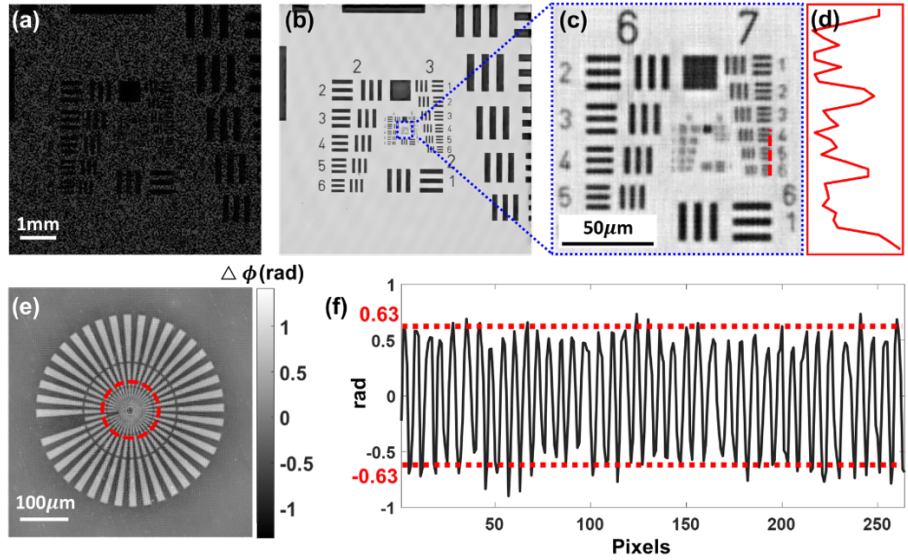

Fig. 5. PS-PtychoLM imaging performance evaluation. (a-d) USAF 1951 resolution target imaging results. (a) Representative PS-PtychoLM raw image of the target. (b) Reconstructed USAF 1951 resolution target image. (c) Enlarged image of the area marked with a blue box in (b). (d) Line profile along dashed red line in (c). (e,f) Quantitative phase target imaging result. (e) Reconstructed phase image. (f) Phase profile along the dashed red circle in (e).

We first imaged a non-birefringent USAF 1951 resolution target (Edmund optics, #64-862) to evaluate the imaging performance of the PS-PtychoLM. An exemplary raw image captured by the PS-PtychoLM is shown in Figure 5a. After the ptychographic reconstruction with 81 image acquisitions, an amplitude image of the resolution target over the entire FoV of 7.07mm x 8.45mm could be reconstructed (Figure 5b). One can see that features in the resolution target can be clearly visualized. Figure 5c is an enlarged image of the area marked with a blue box in Figure 5b. The element 5 in group 7 (line width = 2.46 μm) could be distinguished. The line profile along the dashed red line in Figure 5c is shown in Figure 5d. Our PS-PtyhcoLM prototype achieved a half-pitch resolution of 2.46 μm for a FoV of 7.07 mm x 8.45 mm, which corresponds to an SBP of 9.9 megapixels.

The phase reconstruction performance was then evaluated using a quantitative phase target (Benchmark, Quantitative phase target). Figure 5e depicts the reconstruction result of the phase target using the PS-PtychoLM platform. The phase distribution along the dashed red circle in Figure 5e is presented in Figure 5f. The target features in the reconstructed area shown in Figure 5e are characterized by a height of 200 nm and a refractive index of 1.52. The theoretical phase value, computed from this information and the LED center wavelength (514.5 nm), is indicated by a red dashed line in figure 5f. The reconstructed phase value agrees with the actual value with an error of 6.15%.

### 3.2.2. Birefringent resolution target and depolarizer

To validate the birefringence imaging capability of the PS-PtychoLM, we imaged and quantitatively mapped the phase retardation and optic-axis orientation of the birefringent resolution target (Thorlabs, R2L2S1B). The target consists of liquid crystal polymers between isotropic glass plates and exhibits a uniform phase retardance magnitude across the entire target (provided by the manufacturer). Previous imaging studies on the same target have revealed that the optic-axis orientation difference between the patterned and background regions is approximately 30° [30,44]. Figure 6a,b show the PS-PtychoLM imaging results of the birefringent resolution target. We obtained the phase retardation of 0.15±0.02 rad (mean ± s.d.) across the entire FoV. For the optic-axis orientation, the orientation angles of patterned and background regions were measured to be 155.33°±1.57° and 127.79°±3.72°, respectively. The reconstructed phase retardation values agree with the resolution target specification, in that a uniform phase retardation was measured. In addition, the optic-axis orientation difference between the patterned and background regions was measured to be 27.5°, which matched well with the results from the previous studies. Figure 6c presents birefringence map of the birefringent resolution target, which encodes phase retardation and optic-axis orientation through intensity and color, respectively.

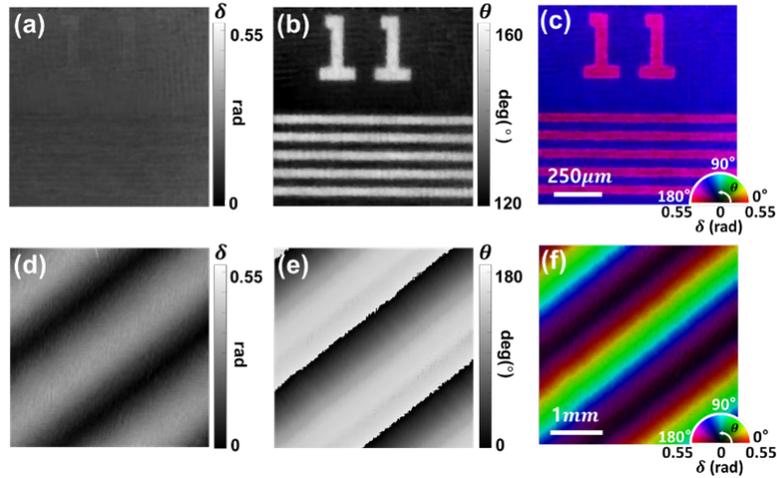

Fig. 6. PS-PtychoLM imaging results of a birefringent resolution target (Thorlabs, R2L2S1B) (a-c) and liquid crystal polymer depolarizer (Thorlabs, DPP25-A) (d-f). (a, d) Phase retardation, (b, e) optic-axis orientation, (c, f) quantitative birefringence map of birefringent resolution target and depolarizer, respectively.

We then performed imaging of a liquid crystal polymer depolarizer (Thorlabs, DPP25-A). This depolarizer consists of a thin film of liquid crystal polymer sandwiched between two glass plates and is designed to have a linearly-ramping phase retardation and optic-axis orientation to convert linearly polarized light into pseudo-randomly polarized light. Figure 6d,e present the PS-PtychoLM imaging results of the depolarizer. Both phase retardation and optic-axis orientation-reconstruction results exhibited spatially periodic variation, in agreement with the information provided by the manufacturer. Figure 6f presents the pseudo-colored birefringence map of the liquid crystal polymer depolarizer.

### 3.2.3. Monosodium urate crystals

We further imaged monosodium urate (MSU) crystals (InvivoGen Co., tlrl-MSU) immersed in phosphate-buffered saline to demonstrate its accuracy in optic axis determination. The MSU crystals are needle-shaped birefringent crystals with strong negative birefringence, i.e., the fast axis is oriented along the axial direction of the crystals [33]. Following the PS-PtychoLM reconstruction, we selected and analyzed a single MSU crystal to verify optic-axis orientation measurements. Figure 7a–h presents representative birefringence maps of the single MSU crystal at various rotation angles. Figure 7i shows a comparison of the optic-axis orientation measured with PS-PtychoLM and that obtained manually with the NIH ImageJ program. The correlation coefficient was measured to be ~0.986, and the phase retardation was found to be ~0.27 rad, regardless of the rotation angle. The standard deviation is indicated by the error bars.

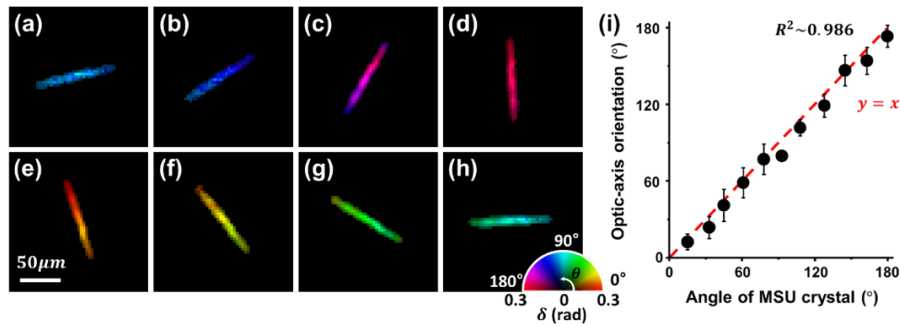

Fig. 7. Validation of PS-PtychoLM optic-axis orientation measurements. (a)–(h) Birefringence maps of a single MSU crystal at different rotation angles (15°,33°,61°,93°,108°,128°,145°, and 178°, respectively). (i) Optic-axis orientation of the MSU crystal measured through rotation at approximately 15° intervals. Remarkable correspondence can be observed between PS-PtychoLM and manual assessments, with $R^2$ = ~0.986.

### 3.2.4. Biological tissue samples

In order to demonstrate PS-PtychoLM imaging capability for biological specimens, we further performed PS-PtychoLM imaging of mouse eye and heart tissues. Paraffin-embedded mouse eye and heart tissue sections were obtained from Asan Medical Center and Yonsei Severance Hospital, Seoul, Republic of Korea, respectively. All experimental procedures were carried out in accordance with the guidelines of the Institutional Animal Care and Use Committees which were approved by Asan Institute for Life Science and Yonsei University College of Medicine. We de-paraffinized the tissue sections (the eye and heart were sectioned at 10 μm and 5 μm thick, respectively) using hot xylene, placed them onto a microscope glass slide, and performed PS-PtychoLM imaging. Shown in Figure 8a,b are the large-FoV quantitative birefringence and phase images of the mouse eye section, respectively. Some parts of the eye, including the cornea and sclera, are formed of a stack of orientated collagen fibers, which result in birefringence with its optic-axis orientated along the fibers [45,46]. Figure 8c-f show the magnified birefringence and phase images of the parts marked with red and orange boxes in Figure 8a, which correspond to the cornea and sclera, respectively. Visualization of the optic-axis orientation was enhanced by overlaying short white lines that indicate the mean optic-axis orientation evaluated over a small region (70 μm × 70 μm). Through the ptychographic phase retrieval algorithm, PS-PtychoLM provides complex and birefringence information of the specimen, which can be used to measure various features of the objects. The phase information provides optical path-length distribution of the sample, while the birefringence map allows for interrogating phase retardation and optic-axis orientation. Figure 8g,h are the images captured with a conventional microscope (Nikon, ECLIPSE Ti, 0.2 NA) after the samples were stained with hematoxylin and eosin (H&E). It is clearly observed that the fiber arrangement observed in Fig. 8g,h greatly matched the optic-axis orientation obtained with PS-PtychoLM.

Figure 8i,j present the birefringence and phase images of the mouse heart section, along with the enlarged images (Figure 8k,l for birefringence and Figure 8m,n for phase) of the parts marked with red and orange boxes in Figure 8i. Because fibrous tissues found in the myocardium exhibit optical birefringence, the optic-axis information may be used to infer myocardium orientation [47]. Figure 8o,p are the corresponding H&E images obtained from the optical microscope (Nikon, ECLIPSE Ti, 0.2 NA); these images correspond to the same area depicted in Figures 8k and l, respectively. The alignment direction of the myocardium observed in the H&E images is consistent with the reconstructed optic-axis orientation in the PS-PtychoLM images. Myofiber disorganization compromises normal heart function and is associated with various cardiovascular diseases such as myocardial infarction [48]. These PS-results suggest that, with further improvement, PS-PtychoLM may serve as a viable PS imaging tool for various biological research and disease diagnosis.

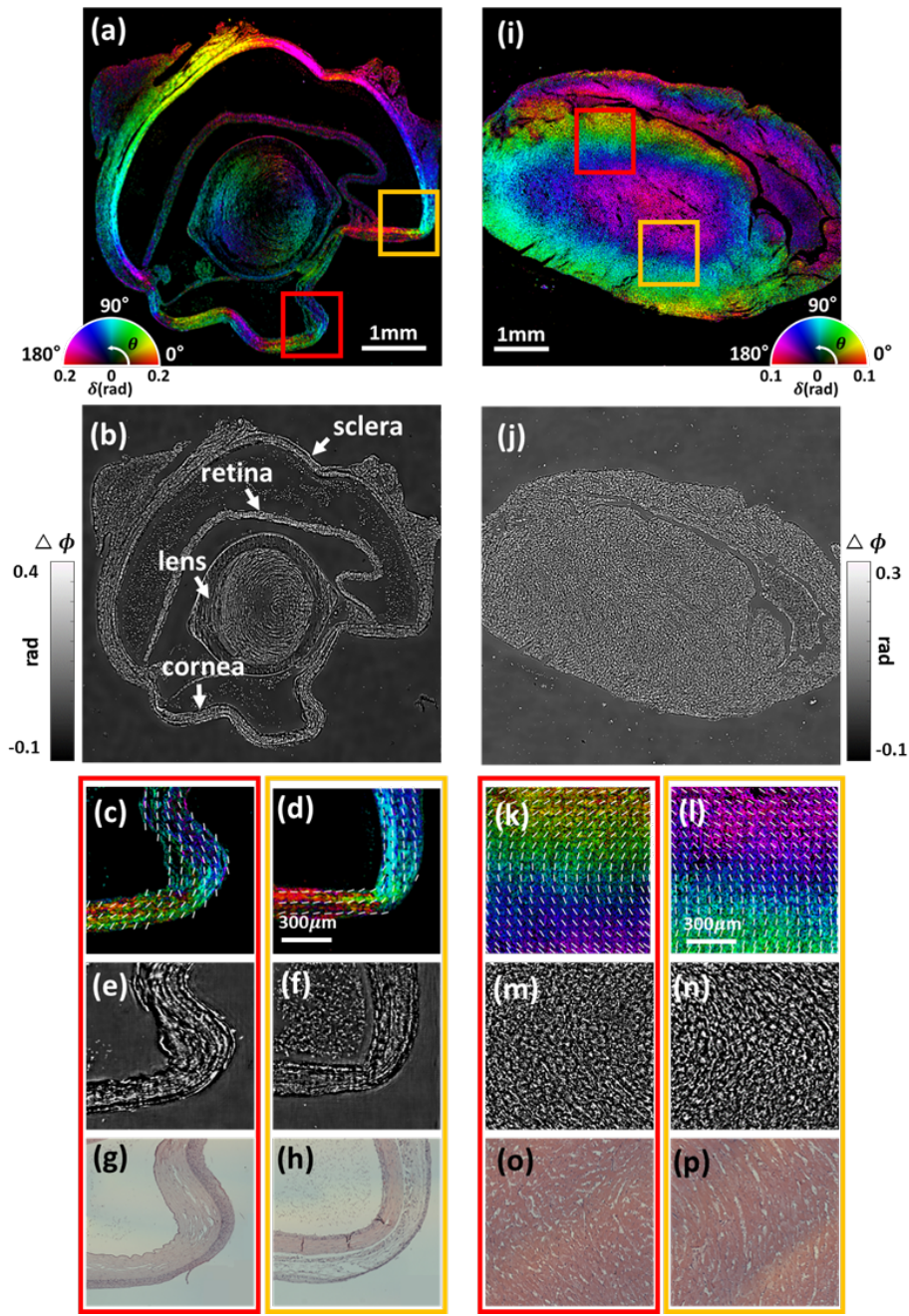

Fig.8. PS-PtychoLM imaging results of mouse eye (a–h) and heart tissue specimens (i–p). (a,i) Quantitative birefringence maps of the mouse eye and heart tissues over the entire FoV. (b,j) Quantitative phase images of the mouse eye and heart tissues over the entire FoV. (c–f) Enlarged birefringence and phase images of the parts marked with red and orange boxes in (a), respectively. (g, h) Hematoxylin and eosin (H&E) staining images of the parts (c, d) captured with an optical microscope (Nikon, ECLIPSE Ti, 0.2NA). (k–n) Enlarged birefringence and phase images of the parts marked with red and orange boxes in (i), respectively. (o, p) H&E staining images of the parts (k, l) captured with an optical microscope (Nikon, ECLIPSE Ti, 0.2NA). Overlaid white short lines on (c, d, k, and l) indicate the mean optic-axis orientation evaluated over a small region (70 μm × 70 μm).

## 4. Conclusion

In summary, we described a computational lens-less birefringence microscope by combining the inertia-free, high-resolution, large-area imaging capability of the mask-modulated ptychography with single-input-state polarization-sensitive imaging setup. The proposed PS-PtychoLM could produce complex-valued information on each polarization channel of the image sensor, and the information was further used to obtain a 2D birefringence map over the entire sensor surface. Our method featured a half-pitch resolution of 2.46 μm across an FoV of 7.07 mm x 8.45 mm, which corresponds to 9.9 megapixels in SBP.

Several features should be noted in the reported platform. We employed the mask-modulated ptychographic configuration with angle-varied LED illumination to achieve inertia-free ptychographic birefringence imaging. The programmable LED array enabled simple, cost-effective, and motion-free operation of PS-PtychoLM, but its short coherence both in spatial and temporal domains compromised the spatial resolution achievable in our method. We used a bandpass filter in the illumination path to improve the temporal coherence, but the spatial coherence could not be improved because either the use of smaller LEDs or larger distance between the LEDs and object compromised the light throughput. Instead of LEDs, coherence light sources such as laser or laser diodes can certainly be employed. These light sources provide higher spatial and temporal coherence, and thus higher spatial resolution may be readily attained. In such cases, however, either the light source or object must be translated to obtain measurement diversity, which is required for ptychographic phase retrieval. We aimed to demonstrate a cost-effective and robust imaging platform and thus employed the LED array as the illumination source instead of the aforementioned solution.

In our PS-PtychoLM implementation, we employed binary mask-assisted ptychographic phase retrieval to obtain complex information of the object. Instead of the binary mask, several recent studies reported on using other random structures such as diffusers with micro-/nano-features to achieve superior resolution [8,17]. However, we note that with the random diffuser in our setup, the object information could not be accurately obtained. Our prototype employed a pixelated polarization image sensor, and it is thus required to decompose the captured image into the ones in different polarization channels and perform pixel interpolation through demosaicing methods. However, the random diffuser significantly scrambles the propagation wavefront of the object wave, and thus information in the missing pixels could not be correctly interpolated by the demosaicing schemes. We numerically investigated the mean squared errors (MSEs) between the information acquired with a full pixel-resolution camera (i.e., without pixel demosaicing method) and the one obtained with pixelated polarization camera and pixel demosaicing. The measured MSE with the random diffuser was found to be ~3x larger than that with the binary mask. Detailed simulation results of PS-PtychoLM imaging with diffuser and binary mask are provided in the Appendix B. We also experimentally validated that the use of a binary mask provided an accurate estimation of polarization information in all the detection channels, thus resulting in accurate birefringence maps. This problem can certainly be alleviated by using a non-polarized full pixel-resolution camera, but it would require mechanical rotation of polarizer/analyzer to reconstruct birefringence maps.

In terms of the mask pattern, we utilized a binary amplitude mask with 50% open channels. Optimal binary mask designs for rapid and robust phase recovery are certainly the subject of future research, as also noted by Ref 36. The influence of the mask design parameters (e.g., feature size and distribution) on the image recovery performance is being investigated. The results are expected to provide useful guidelines for binary mask designs in the mask-assisted ptychographic imaging systems.

We expect that our method would have a broad range of applications in material inspections and quantitative biological studies. For example, large-FoV, high-resolution birefringence imaging can be used to inspect internal features (e.g., molecular orientation of the liquid crystal layer between two substrates) of liquid crystal-based devices [49] and to detect defects of semiconductor wafers in the manufacturing process [23]. In addition, birefringence imaging can be used in several biomedical applications such as for malaria detection [50,51], brain slide imaging [52], retina imaging [53,54], cancerous cell differentiation [55,56], and bulk tissue characterization [57]. On a technical note, PS-PtychoLM can be extended to lens-less birefringence tomography by combining multi-slice beam propagation (MBP) forward model for 3D structures [58,59] and vectorial Jones-based analysis [44,60,61]. Our demonstration is limited to imaging only thin and transparent samples. However, implementation of the MBP model on the lens-less imaging platform would enable imaging of thick multiple scattering samples through a combination with illumination engineering methods (e.g., angle scanning).

## Appendix A: Comparison of various demosaicing methods

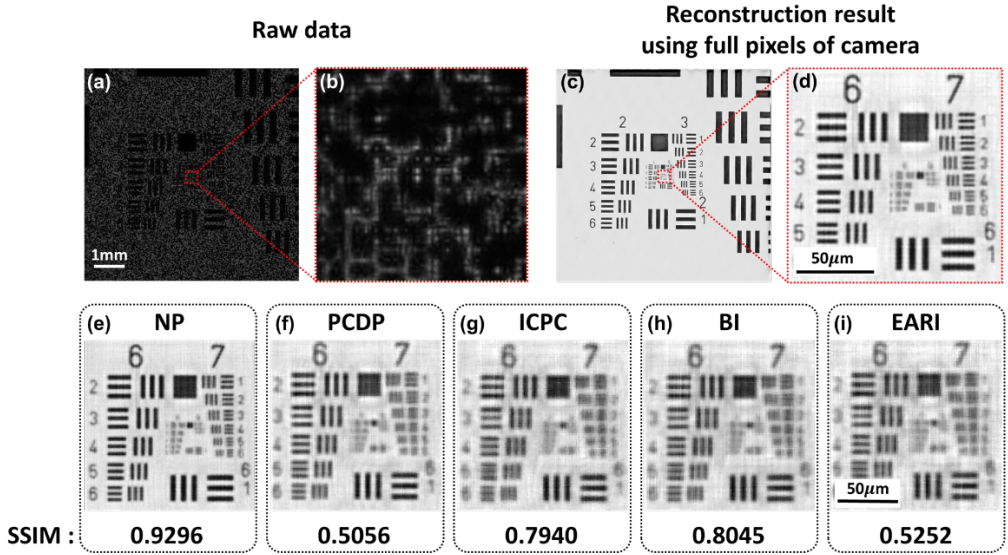

Fig. 9. Compared image reconstruction performances of various pixel demosaicing methods. (a) Representative raw image of an USAF resolution target captured with an LED along the optical axis. (b) Enlarged image of the area marked with a red box in (a). (c) Reconstructed image of the target using the full pixels of the camera (i.e., without pixel demosaicing method). (d) Enlarged image of the area marked with a red box in (c). (e-i) Reconstruction results with various demosaicing methods. The acquired image was decomposed into the ones of each polarization channel and missing pixel information was estimated through (e) Newton's polynomial (NP), (f) polarization channel difference prior (PCDP), (g) intensity correlation among polarization channels (ICPC), (h) bilinear interpolation (BI), and (i) edge-aware residual interpolation (EARI) algorithms, respectively. The structural similarity index measure (SSIM) values for each reconstruction result were evaluated against the result in (c) (i.e., the result without pixel demosaicing) and are presented.

In order to find a suitable demosaicing method for PS-PtychoLM, we evaluated various pixel demosaicing methods including Newton's polynomial (NP) [43], polarization channel difference prior (PCDP) [62], intensity correlation among polarization channels (ICPC) [63], bilinear interpolation (BI), and edge-aware residual interpolation (EARI) algorithms [64]. We performed PS-PtychoLM imaging of a USAF resolution target and compared the reconstruction performances with the aforementioned demosaicing methods. A representative raw intensity image captured using the PS-PtychoLM is shown in Figure 9a, along with an enlarged image of the area marked with a red box in Figure 9a (Figure 9b). As can be noted, the features in the object could not clearly be discerned because the object information was largely obscured by the mask-modulated object wave. Upon the acquisition of 81 images with angle-varied LED illuminations, we separated each image into the ones in four polarization channels and the missing pixel information in each channel was estimated using the interpolation methods. We also carried out image reconstruction using the full pixel information of the camera without demosaicing as the reference. Figure 9c,d present the image reconstruction results obtained using the full pixel information of the image sensor. It is evident that, after ptychographic retrieval, the features in the target can be clearly visualized. The reconstruction results using NP, PCDP, ICPC, BI, and EARI interpolation algorithms are presented in Figure 9e-i, along with their structural similarity index measure (SSIM) values computed with reference to the result in Figure 9c. It can be noted that the NP interpolation method provided superior image reconstruction compared to other methods. This superior performance of NP compared with others may be accounted for by that the NP operates on the polynomial interpolation error estimation, thus being much effective in preserving both low- and high-spatial-frequency information [43].

## Appendix B: Birefringence reconstruction results with random diffuser and binary mask

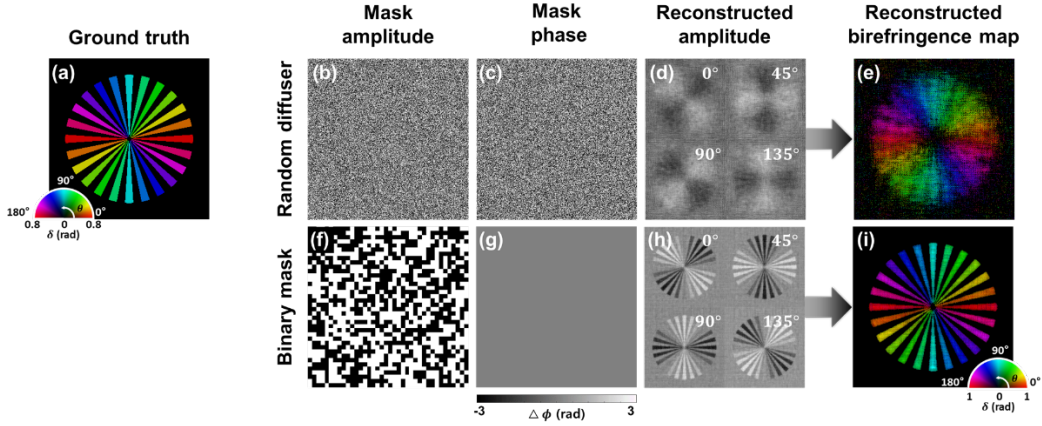

Fig. 10. Comparison of the PS-PtychoLM birefringence reconstruction performance with an unknown random diffuser and pre-defined binary mask. (a) Ground truth birefringence information of a simulation target. (b,f) Amplitude and (c,g) phase information of the random diffuser and binary mask, respectively. (d,h) Reconstructed amplitude information of each polarization channel and (e,i) obtained birefringence map using the random diffuser and binary mask, respectively.

We employed a pre-defined binary structure as the modulation mask in our PS-PtychoLM implementation. This mask served as a constraint in the iterative phase retrieval process, which helps, along with the captured images, convergence to the solution. We separately evaluated a random mask with unknown micro-/nano- features to examine if it can be used in our PS-PtychoLM imaging. The random mask has been employed in various forms of lens-less ptychographic microscopes to achieve higher spatial resolution [8,17]. Our results are presented in Figure 10. We considered a birefringence object depicted in Figure 10a as the target and performed numerical simulations with an unknown random diffuser and binary mask. Figure 10b,c present the amplitude and phase distributions of the random diffuser used in the simulation. The feature size of the diffuser was set to be 3.45 μm, and its amplitude and phase values were uniformly distributed in the interval [0, 1] and [-π, π], respectively. The reconstructed amplitude images of each polarization channel and birefringence information are shown in Figure 10d,e. The amplitude information of the target could not be reconstructed accurately, and consequently, the reconstructed birefringence map also significantly deviated from the ground truth. This can be explained in part by that the random diffuser significantly scrambles the propagation wavefront of the object exit wave, and thus polarization states in the missing pixels could not be correctly interpolated by Newton's polynomial demosaicing method [43]. Note that we employed Newton's polynomial demosaicing method due to its superior interpolation performance compared with others. (see Appendix A).

In contrast, the use of a pre-determined binary mask with a mean feature size of 27.6 μm enabled an accurate reconstruction of the object birefringence information. Figure 10f,g are the amplitude and phase of the binary mask used in the simulation, respectively. The amplitude of the mask consisted only of 0 and 1, and the phase was constant across the plane (phase = 0 in this simulation). The reconstructed amplitude images of each polarization channel and birefringence map clearly demonstrate that the binary mask enabled accurate estimation of the polarization information throughout all polarization channels, thus resulting in accurate birefringence map.

**Funding** Korea Medical Device Development Fund (KMDF_PR_20200901_0099); Commercialization Promotion Agency for R&D Outcomes(COMPA) ((2022)FLRECTD04_2-3); Samsung Research Funding and Incubation Center of Samsung Electronics (SRFC-IT2002-07); National Research Foundation of Korea (2020R1A2C201206111).

**Disclosures.** The authors declare no conflicts of interest.

**Data availability.** Data underlying the results presented in this paper are not publicly available at this time but may be obtained from the authors upon reasonable request.